\documentclass[a4paper, conference]{IEEEtran}
\IEEEoverridecommandlockouts
\usepackage{cite}
\usepackage{amsmath,amssymb,amsfonts}
\usepackage{algorithmic}
\usepackage{graphicx}
\usepackage{textcomp}
\usepackage{subfigure}
\usepackage{graphicx}
\usepackage{wrapfig}
\def\BibTeX{{\rm B\kern-.05em{\sc i\kern-.025em b}\kern-.08em
    T\kern-.1667em\lower.7ex\hbox{E}\kern-.125emX}}
\begin{document}

\title{Hole Mobility Model for Si
Double-Gate Junctionless Transistors
}

\author{\IEEEauthorblockN{Fan Chen}
\IEEEauthorblockA{\textit{Network for Computational }\\
\textit{Nanotechnology (NCN) }\\
Purdue University \\
West Lafayette, IN 47906\\
fanchen.purdue@gmail.com}
\and

\IEEEauthorblockN{Kangliang Wei}
\IEEEauthorblockA{\textit{Institute of Microelectronics} \\
\textit{Peking University}\\
Beijing 100871 \\
gary.weikangliang@gmail.com}
\and

\IEEEauthorblockN{Wei E. I. Sha}
\IEEEauthorblockA{\textit{College of Information Science }\\
\textit{ \& Electronic Engineering}\\
\textit{Zhejiang University}\\
Hangzhou 310027, China\\
weisha@zju.edu.cn}

\and
\IEEEauthorblockN{Jun Z. Huang}
\IEEEauthorblockA{\textit{Network for Computational }\\
\textit{Nanotechnology (NCN) }\\
\textit{Purdue University}\\
West Lafayette, IN 47906\\
junhuang1021@gmail.com}
}

\maketitle

\begin{abstract}
In this work, a physics based model is developed to calculate the hole mobility of ultra-thin-body double-gate junctionless transistors. Six-band $k\cdot p$ Schr\"{o}dinger equation and Poisson equation are solved self-consistently. The obtained wave-functions and energies are stored in look-up tables. Hole mobility can be derived using the Kubo-Greenwood formula accounting for impurity, acoustic and optical phonon, and surface roughness scattering. Initial benchmark results are shown.
\end{abstract}

\vspace{-4mm}
\section{Introduction}
\vspace{-4mm}
The continuous scaling of MOSFETs has met tremendous challenges such as large leakage current, high power consumption. Novel material or transistor design have been shown to have promises in overcoming these challenges \cite{chen2016configurable, chen2017thickness, tan2016first, chen2017transport, chen2017switching, fay2017iii}. However, much more progress are still needed to realize large-scale manufacturing in such proposals \cite{chen2016transport, chen2015transport}. Junctionless transistors (JLT), on the other hand, use small doping gradient between the silicon based channel and contacts to avoid ultra-shallow source/drain junction. Besides, it is also shown that the JLT has lower gate capacitance and off-state leakage current than conventional MOSFETs \cite{lee2009junctionless}, making it a very promising candidate for future nanoscale devices. The high doping in JLT was considered to degrade the mobility by strong ionized impurity scattering \cite{lee2009junctionless}. However, this mobility degradation can be largely reduced by the screening effect from the free carriers, especially for ultra-thin bodies \cite{goto2012mobility}. The simulation work for electrons has been done that helps better understanding of this process \cite{wei2014physically}. As their indispensability for future CMOS technology, it is meaningful to investigate the hole mobility in p-type double-gate JLTs. In contrast to the effective mass electron mobility model in \cite{wei2014physically}, six-band $k\cdot p$ needed to be employed here to account for the heavy, light and split-off sub-bands.
In this paper, we provide a detailed method of physics-based modeling of hole mobility in UTB Si devices and some initial results. \\

\section{Methods}
\subsection{Six-band $k\cdot p$ Schrodinger-Poisson Self-consistent Calculation}
Unlike the conduction band structure in silicon, the valence band at the symmetry point $\Gamma$ is more complicated due to the coupling of the heavy, light, and split-off hole bands. Thus, at least six bands should be considered to calculate the mobility in silicon based pMOSFETs. A six-band $k\cdot p$ method is employed for bandstructure and wavefunction calculation \cite{fischetti2003six}. The six-band $k\cdot p$ Schr\"{o}dinger equation can be described using the following formula: 

\begin{equation}
\left(H\left(\textbf{k},k_y\right)+u\left(y\right)\cdot \textbf{I}\right)\cdot \psi_\textbf{k}\left(y\right) = E\left(\textbf{k}\right) \psi_\textbf{k}\left(y\right) \label{scheq}
\end{equation}
in which the confinement direction is $y$, \textbf{k} = $\left(k_x, k_z\right)$ is a two-dimensional in-plane wave vector. $\psi_\textbf{k}\left(y\right)$ a six dimension wave vector that satisfies to the boundary condition $\psi_\textbf{k}\left(0\right) = \psi_\textbf{k}\left(T\right) = 0$ with $T$ being the thickness of the UTB. Due to the confinement in $y$ direction, $k_y$ can be substituted by $-id/dy$. Then \eqref{scheq} is solved numerically for each in-plane grid $\textbf{k}=(k_x, k_z)$. With a mesh size of $N_y$ along $y$ direction, the final matrix is in a tridiagonal block form with the size of $6N_y\times6N_y$. In the six-band $k\cdot p$ Hamiltonian, spin-orbital coupling is considered and the parameters are taken from \cite{moussavou2015influence}.
The potential $u\left(y\right)$ is obtained by Poisson equation with the same mesh size $N_y$:
\begin{equation}
\frac{d}{dy}\epsilon\left(y\right)\frac{d}{dy}u\left(y\right)= -e^2\left(\rho_h\left(y\right)+N_D\left(y\right)\right) \label{poieq}
\end{equation}
in which $\epsilon$ is the permittivity, $N_D\left(y\right)$ is the concentration of ionized donors. $\rho_h\left(y\right)$ = $\Sigma_{occupied \ \textbf{K}} \ \psi^{\dagger}_{\textbf{k}}\left(y\right)\cdot\psi_{\textbf{k}}\left(y\right)$ is the hole density obtained from \eqref{scheq}. This Hole density is then passed to \eqref{poieq} to update the electrostatic potential. The calculation is iterated until the final convergence criteria is reached. After this $k\cdot p$ Schr\"{o}dinger-Poisson self-consistent calculation, the final eigenstates for each $\left(E,\textbf{k}\right)$ point will be stored in a look-up table to evaluate mobility. For each sub-band $v$, the density of states at a given energy $E$ is given by: \\

\begin{equation}
\rho_v\left(E\right) = \theta\left[E-E^{\left(0\right)}_v\right]\frac{2}{(2\pi)^2} \int^{\pi}_0d\phi \frac{k_v\left(E,\phi\right)}{|\frac{\partial E_v}{\partial k}|_{k_v\left(E,\phi\right)}}\label{dosequ}
\end{equation}
where $E_v\left(0\right)$ is the $v$th subband energy at $\Gamma$, $\theta\left(x\right)$ is the step function. The $\textbf{k}$ in \eqref{scheq} can be projected in to polar coordinates by $\left(k_x,k_z\right) = \left(k\cos\phi, k\sin\phi\right)$. This polar angle $\phi$ is divided into $N_{\phi}-1$ intervals. In each interval of $\phi$, the $k$ values in the look-up table with energy values within $E+\delta E$  is the $k_v\left(E,\phi\right)$ in \eqref{dosequ}. Here $\delta E = |E_{max}-E_{min}|/N_E$. After all, the sum of the density of states of each subband $v$ times the Fermi-Dirac function gives the total hole density:
\begin{equation}
\begin{aligned}\label{den}
\rho(y) &= \sum_{v} \int d\textbf{k} \cdot f\left(E_v\left(\textbf{k}\right)-E_f\right)\cdot |\psi^v\left(y\right)|^2 \\
 &= \sum_{v}  |\psi^v_0\left(y\right)|^2 \int_{E_{min}}^{E_{max}} dE \cdot \frac{\rho_v\left(E\right)}{1+\exp{\left(\frac{E-E_f}{k_BT}\right)}}
\end{aligned}
\end{equation}
The wavefunction $\psi^v\left(y\right)$ at each $E\left(\textbf{k}\right)$ is approximated by the wavefunction at $\Gamma$ point $\psi^v_0\left(y\right)$ to save the total simulation time. Here, $f$ is the Fermi-Dirac function and $k_B$ is the Boltzman constant.

\subsection{Physics Based  Hole Mobility Model}

The matrix elements and the scattering rates are obtained from Fermi's golden rule. The momentum relaxation time approximation is applied for mobility calculation. The three important scattering mechanisms employed here for junction-less devices are: 1) impurity scattering,  2) phonon scattering including intra- and inter-valley scattering for acustic, optical and polar optical phonons, and 3) surface roughness scattering. Once the scattering rates are known, we can calculate the mobility by applying the Kubo-Greenwood formula.\\
\subsubsection{Impurity Scattering}
Due to the high doping in JLTs, impurity scattering is an important mechanism that affects the mobility. The scattering rate for ionized impurities is obtained by calculating the matrix elements with the screened Coulomb potential produced by point charges located at $y_0$, which can be expressed by the Fourier transformation of the scattering potential\cite{esseni2011nanoscale}.
\begin{equation}
M_{n,n'}^{(0)}(k,k',y_0) = \int_y\psi_{n'k'}^{\dagger}(y)\cdot \psi_{nk}(y)\phi_{pc}(q,y,y_0)dy
\end{equation}
The scattering potential inside the channel can be expressed as: \\
\begin{equation}
\begin{aligned}
&\phi_{pc}(q,y,y_0) = \frac{e}{2q\epsilon_{si}}[e^{-1|y-y_0|} + C_1 e^{qy} + C_2 e^{-qy}] \\
&C_1 = \frac{(\epsilon_{si} - \epsilon_{ox})^2e^{-q|y_0|}+(\epsilon_{si}^2 - \epsilon_{ox}^2)e^{-q|T - y_0| - T}}{(\epsilon_{si} + \epsilon_{ox})^2\cdot e^{2qT} - (\epsilon_{si} - \epsilon_{ox})^2}\\
&C_2 = \frac{(\epsilon_{si} - \epsilon_{ox})(C_1 + e^{-q|y_0|})}{(\epsilon_{si}+\epsilon_{ox})}
\end{aligned}
\end{equation}

Screening by electrons in multiple subbands as well as intra- and inter-subband transitions have been taken into account. The screened matrix element: $M_{v,m,m}^{(scr)} = \frac{M_{v,m,m}(q)}{\epsilon_D \cdot {q}}$, where $\epsilon_D$ is the determinant of dielectric matrix $\epsilon$ and the polarization $\Pi_{n,n'}(q)$  factor are defined as: 
\begin{equation}
\begin{aligned}
&\epsilon_D (\textbf{q}) = 1+\sum_{n}\chi_{w,n,n}(\textbf{q}) = 1-\sum_{n}\frac{e^2}{\textbf{q}(\epsilon_{si}+\epsilon_{ox})} \Pi_{n,n}(\textbf{q})\\
&\Pi_{n,n'}(\textbf{q}) = \frac{1}{A}\sum_{\textbf{k}}\frac{f_{n'}(\textbf{k}+\textbf{q})-f_n(\textbf{k})}{E_{n'}(\textbf{k}+\textbf{q})-E_n(\textbf{k})}
\end{aligned}
\end{equation}
where $f_n(\textbf{k})$ is the occupation function of subband $n$. Finally the impurity scattering relaxation time can be obtained: 
\begin{equation}
\begin{aligned}
\frac{1}{\tau_{IM}^{n'}}(E,\theta) &= \frac{1}{2\pi \hbar} \sum_n \int_0^{2\pi} k_n(E,\theta+\beta)\frac{dk_n(E,\theta+\beta)}{dE}\\
&\cdot |M_{n,n'}|^2 (1-cos\beta)d\beta
\end{aligned}
\end{equation}

\subsubsection{Phonon Scattering}
Elastic, energy equipartition, and isotropic approximation are made for acoustic phonon scattering. $\psi(y)$ is approximated by the wavefunction at $\Gamma$ point $\psi^v_0(y)$ to save the total simulation time. Both the inter and intra band acoustic scatterings are considerred. The momentum relaxation time for acoustic phonon and for sub-band $\mu$ is \cite{chang2015investigation}:
\begin{equation}
\frac{1}{\tau_{ac}^\mu} = \frac{2\pi k_bTD_{ac}^2}{\rho \hbar v_l^2} \sum_v F_{\mu, v} \rho_v (E)\label{acuphequ}
\end{equation}
Here, $\rho$ is the crystal density and $v_l$ is the longitudinal sound velocity. $D_{ac}$ is the effective acoustic deformation potential. The form factor $F_{\mu, v}$ that represents how likely the carrier can be scattered from sub-band $v$ to sub-band $\mu$ can be calculated from the wave-function overlap along the confinement direction:\\
\begin{equation}
 F_{\mu, v} = \int _0^T |\psi_0^{v\dagger}(y)\cdot \psi_0^\mu(y)|^2 dy\label{formequ}
\end{equation}
Similarly, the momentum relaxation rate with regard to the optical phonon scattering can be expressed as:\\
\begin{equation}
\begin{aligned}
\frac{1}{\tau_{op}^\mu} &= \frac{\pi D_{op}^2}{\rho \omega_{op}} \sum_v F_{\mu, v} \cdot \left(n_{op}(\omega_{op})+\frac{1}{2}\mp \frac{1}{2}\right)\\
&\cdot \frac{1-f_0(E\pm \hbar \omega_{op})}{1-f_0(E)} \cdot \rho_v (E \pm \hbar \omega_{op})\label{opphequ}
\end{aligned}
\end{equation}
Here, $n_{op} = 1/\left(\exp(\hbar \omega_{op}/k_BT) - 1\right)$ is the Bose-Einstein Distribution which gives the number of the optical phonons with energy $\hbar \omega_{op}$. The plus and minus sign represents the absorption and  emission process respectively. $D_{op}$ is the optical phonon deformation potential. Polar optical phonon scattering is suggested to be important for III-IV materials and is not considered in this work for Si based transistors.\\ 
\subsubsection{Surface Roughness Scattering}
The scattering matrix element from sub-band $\mu$ to sub-band $v$ is given by calculating the derivative of the wavefunction function at the interface. For double-gate structure, both the interfaces at $y = 0$ and $y=T$ need to considered \cite{chang2015investigation}: \\
\begin{equation}
\Gamma^{SR}_{v,\mu} = -\left(\frac{\psi_v^{\dagger}(0)}{dy}\cdot A \cdot \frac{\psi_u(0)}{dy} + \frac{\psi_v^{\dagger}(T)}{dy}\cdot A \cdot \frac{\psi_u(T)}{dy}\right)
\end{equation}
$A$ can be derived from the equation $H_0 = A\frac{d^2}{dy^2} + B +e\phi(y)\textbf{I}$ \cite {chang2015investigation}. Surface roughness power spectrum $S(q) = \frac{\pi\Delta^2 \Lambda^2}{(1+q^2\Lambda^2/2)^3}$, in which $\textbf{q} = \textbf{k}_v-\textbf{k}_\mu$ is the change of the $\textbf{k}$ vector, $\Delta$ is the height and $\Lambda$ is the correlation length. The scattering potential $V_{vk_v,\mu k_\mu}^{SR} = S(\textbf{q})\cdot \Gamma^{SR}_{v,\mu}$. Finally the relaxation time of surface roughness scattering can be obtained by:\\
\begin{equation}
\begin{aligned}
\frac{1}{\tau_{SR}^\mu}(E,\theta) &= \frac{1}{2\pi \hbar} \sum_v \int_0^{2\pi} k_v(E,\theta+\beta)\frac{dk_v(E,\theta+\beta)}{dE}\\
&\cdot |V^{SR}_{vk_v,\mu k_\mu}|^2 (1-\cos\beta)d\beta
\end{aligned}
\end{equation}
In which, $\beta$ is the angle between initial state $\textbf{k}_\mu$ and final state $\textbf{k}_v$. Both intra and inter sub-band surface roughness scattering are considered in the same manner. 
\section{Results And Disucssion}\label{AA}
\begin{figure}[!t]
\centering
\subfigure[]{\includegraphics[width=0.24\textwidth]{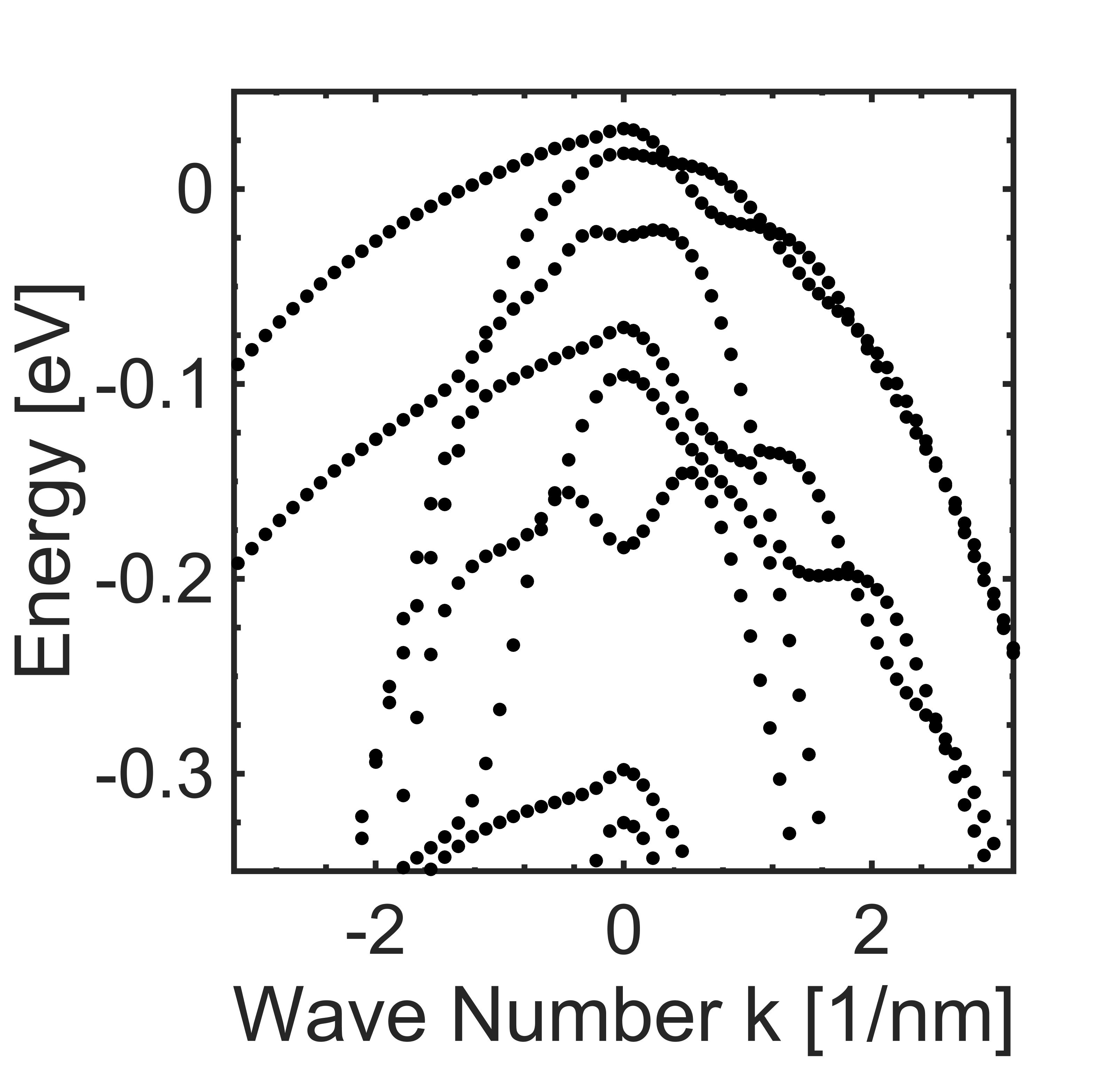}%
\label{2a}}
\hfil
\subfigure[]{\includegraphics[width=0.24\textwidth]{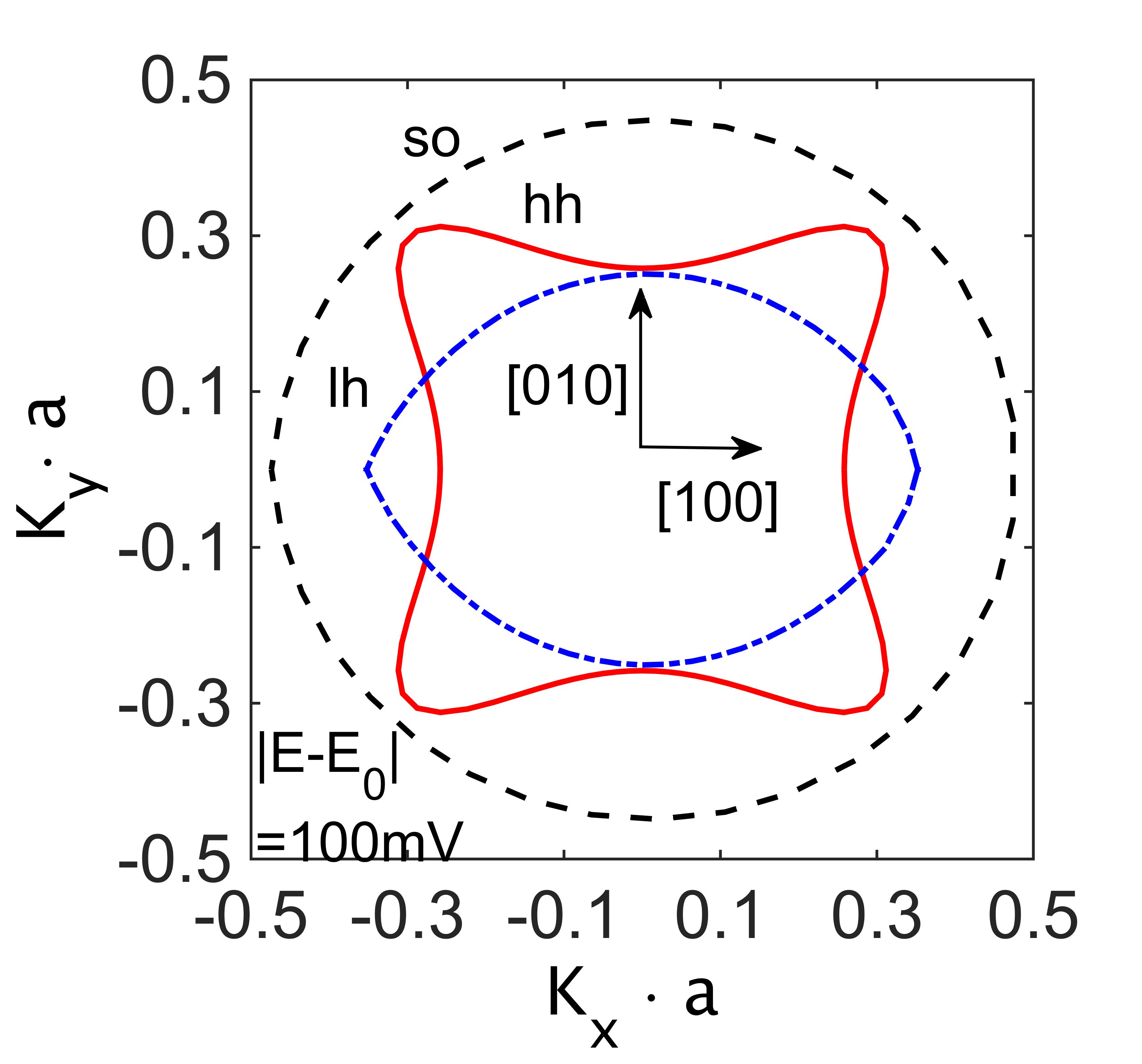}%
\label{2b}}
\vfil
\subfigure[]{\includegraphics[width=0.24\textwidth]{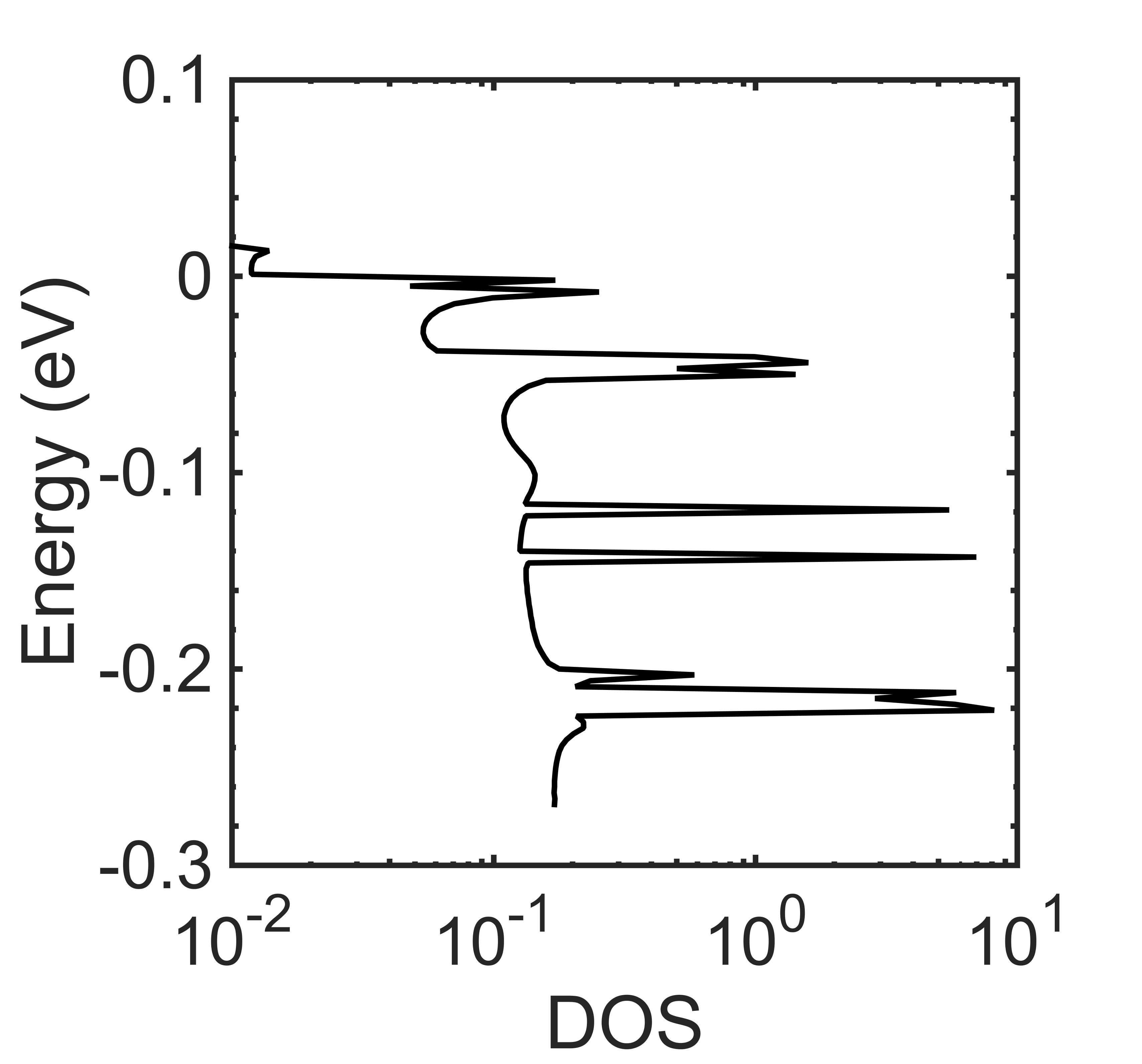}%
\label{2c}}
\hfil
\subfigure[]{\includegraphics[width=0.24\textwidth]{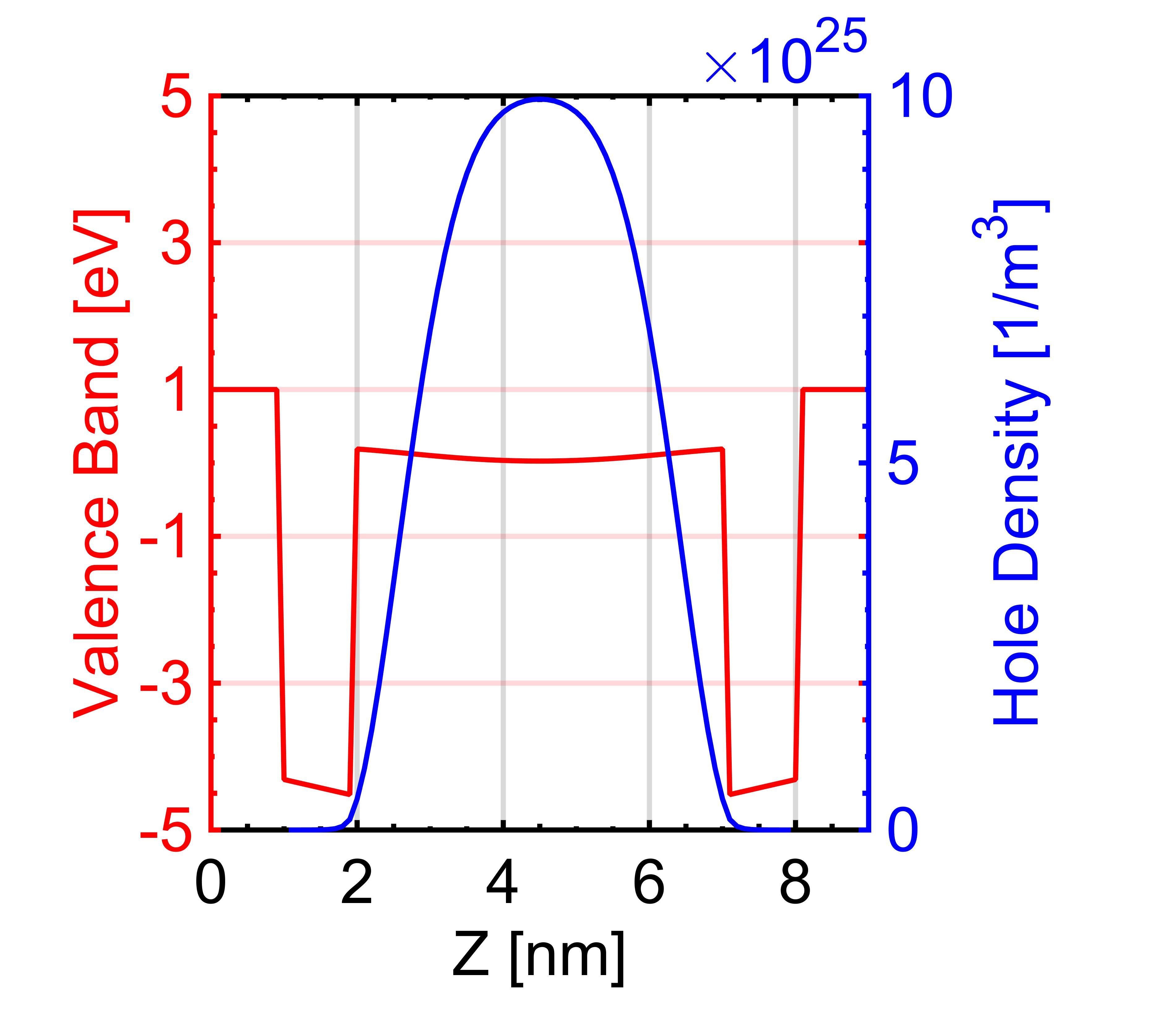}%
\label{2d}}
\caption{(a) E-K diagram of silicon valence bands from six-band $k\cdot p$ Hamiltonian (b) The equal-energy lines with energy -70meV. The heavy hole, light hole and split-off sub-bands are present at this energy level. (c) Density of sates plot with respect to energy. The spikes show the maximum of one sub-band. (d) Potential profile and hole distribution along the confinement direction.}
\label{22}
\end{figure}

Some initial band profile and benchmark results will be discussed in this section. The ultra-thin-body double-gate JLTs studied in this work has a doping concentration of $N_a = 1\times10^{20}cm^{-3}$, channel thickness $T=5nm$, and oxide thickness of 1nm. The devices considered here are long-channel devices. Fig.~\ref{22} (a) shows the energy dispersion from $k\cdot p$ matrix. An energy range of $300meV$ is sufficient to cover the Fermi tail at room temperature. As shown in Fig.~\ref{22} (a), from the maximum of the valence band $E_{max} = 0.03eV$ to $E_{min} = -0.27eV$, six sub-bands will be considered in the transport calculation. Fig.~\ref{22} (b) shows the equal-energy lines in the $k_x-k_y$ plane with the energy $100meV$ below $E_{max}$. Only three sub-bands shows up at this energy level, they are the heavy hole, light hole and split-off bands respectively. In Fig.~\ref{22}b, the density of states (DOS) with respect to the energy is shown. Since the DOS is integrated in the $k_x-k_y$ plane, it can be non-integer numbers. The spikes in the plot represents the maximum of one sub-band. The maximum point in E-K diagram usually has a large DOS. Fig.~\ref{22} (d) is the potential profile and the hole density along the confinement direction. Due the double gate structure, symmetric profiles are expected. As we can see, the high density of holes requires a proper treatment of the screening effect and impurity scattering.

\bibliographystyle{IEEEtran}
\bibliography{all}

\end{document}